\documentclass[aps,prb,twocolumn,amsmath,amssymb,showpacs]{revtex4}

\usepackage{graphicx}
\usepackage{dcolumn}
\usepackage{bm}

\begin{document}


\title{Microscopic theory of multipole ordering
  in NpO$_2$
}

\author{Katsunori Kubo}
\author{Takashi Hotta}
\affiliation{%
  Advanced Science Research Center, Japan Atomic Energy Research Institute,
  Tokai, Ibaraki 319-1195, Japan}

\received{18 February 2005}
\published{29 April 2005}

\begin{abstract}
  In order to examine the mysterious ordered phase
  of NpO$_2$
  from a microscopic viewpoint,
  we investigate an $f$-electron model on an fcc lattice
  constructed based on a $j$-$j$ coupling scheme.
  First,
  an effective model with multipole interactions is derived
  in the strong-coupling limit.
  Numerical analysis of the model
  clearly indicates that
  the interactions for $\Gamma_{4u}$ and $\Gamma_{5u}$ moments
  are relevant to the ground state.
  Then, by applying mean-field theory to the simplified model
  including only such interactions,
  we conclude that
  longitudinal triple-$\bm{q}$ $\Gamma_{5u}$ octupole order
  is realized in NpO$_2$
  through the combined effects of multipole interactions
  and anisotropy of the $\Gamma_{5u}$ moment.
\end{abstract}

%

\maketitle

The magnetism of actinide compounds has been one of the fascinating phenomena
in the research field of condensed matter physics,
since we have observed various ordered states
at low temperatures due to the competition among
several kinds of interactions between $5f$ electrons.~\cite{Santini2}
Among them,
the phase transition in NpO$_2$~\cite{Westrum} is peculiar,
and the determination of
the order parameter has been a long-standing issue
for many years.
At first, this transition was expected to have a magnetic origin,
due to the specific heat behavior
similar to that of the antiferromagnet UO$_2$.
In fact,
a cusp in magnetic susceptibility has been also found
at the transition temperature $T_0 \simeq 25$~K,
as in an antiferromagnetic transition.~\cite{Ross,Erdos}
However, no magnetic reflection has been observed
in neutron diffraction experiments~\cite{Heaton}
and more sensitive probes
such as M\"{o}ssbauer~\cite{Dunlap,Friedt}
and $\mu$SR~\cite{Kopmann} measurements
have detected only a very small internal magnetic field.
If it is caused by a dipole moment, the magnitude should be
of the order of 0.01$\mu_{\text{B}}$.~\cite{Dunlap}
Thus, the transition is \textit{not} the usual magnetic one, even though
time reversal symmetry is broken.

Since the possibility of an ordinary magnetic transition is discarded,
it may be natural to consider that the interesting properties of
NpO$_2$ come from the orbital degeneracy of the $f$ electrons.
However, typical orbital order, i.e., quadrupole order,
does not break time reversal symmetry, and thus,
we also discard this possibility.
Finally, one may consider ordering of higher-order multipoles
such as octupoles.~\cite{Santini}
In fact, recent results of resonant X-ray scattering experiments
are consistent with this scenario.~\cite{Mannix,Paixao,Lovesey}
In order to explain superstructure Bragg peaks below $T_0$
due to longitudinal triple-$\bm{q}$ $\Gamma_{5g}$ quadrupole order,
it has been proposed that longitudinal
triple-$\bm{q}$ $\Gamma_{5u}$ octupole order is realized,~\cite{Paixao}
since it induces quadrupole order and
breaks time reversal symmetry.
Very recent experiments on the $^{17}$O NMR also
support the triple-$\bm{q}$ ordered state.~\cite{Tokunaga}

In order to understand such multipole ordering, phenomenological
theories have been developed up to now, assuming the existence of
triple-$\bm{q}$ octupole order.
These theories have consistently explained the experimental facts,
such as a cusp in magnetic susceptibility,
broken time reversal symmetry, no distortion
from cubic structure,~\cite{Mannix}
a quadrupole moment observed in the resonant X-ray scattering,
and the structure of $^{17}$O-NMR spectra.~\cite{Sakai2}
However, the origin of octupole order cannot be discussed
in phenomenological theories.
It was necessary to proceed
to a microscopic theory to understand
\textit{why such higher-order multipole order is realized in NpO$_2$},
but it has been a very hard task
to study multipole ordering from a microscopic level,
since multipole moments originate from the complex combination
of spin and orbital degrees of freedom of $f$ electrons.

In this paper, we attempt to overcome such a difficulty
by considering a microscopic model
based on a $j$-$j$ coupling scheme.~\cite{Hotta}
Following the procedure to estimate the superexchange interaction
in $d$-electron systems, we derive an effective
multipole interaction model from a Hamiltonian
on an fcc lattice corresponding to NpO$_2$.
The correlation functions for multipole moments are numerically
evaluated, suggesting that the interactions for
$\Gamma_{4u}$ and $\Gamma_{5u}$ moments are relevant to the ground state.
Then, we apply a mean-field theory to
a simplified model including only such interactions.
It is shown that the ground state has
the longitudinal triple-$\bm{q}$ $\Gamma_{5u}$
octupole order, proposed phenomenologically
for the low-temperature phase of NpO$_2$.~\cite{Paixao}
We also evaluate specific heat and magnetic susceptibility
for the triple-$\bm{q}$ octupole ordered state.

First we briefly explain our approximation to treat the multipole state.
In general, Coulomb interaction, symbolically expressed by ``$U$'',
is larger than the spin-orbit interaction $\lambda$,
leading to an $LS$ coupling scheme,
but the energy scale of the present problem is $T_0$,
much smaller than both $U$ and $\lambda$.
Since we find that the local ground state does not
qualitatively depend on the order to take infinite limits of
$\lambda/T_0$ and $U/T_0$,
we prefer to use a $j$-$j$ coupling scheme,
in which we accommodate $f$ electrons
among sextet with total angular momentum $j$=5/2.
For actinide dioxides, we propose to ignore further
two states in the sextet.
Due to crystalline electric field (CEF) effect for the
CaF$_2$ structure,
the sextet is split into $\Gamma_8$ quartet and $\Gamma_7$ doublet.
In this case, the $\Gamma_7$ state is higher than the $\Gamma_8$ level
and the splitting energy is defined as $\Delta$.
When we accommodate two, three, and four electrons in the $\Gamma_8$
level, the ground states are $\Gamma_5$, $\Gamma^{(2)}_8$,
and $\Gamma_1$, respectively,~\cite{Hotta2}
consistent with the CEF ground states of UO$_2$,~\cite{Amoretti}
NpO$_2$,~\cite{Fournier} and PuO$_2$,~\cite{Kern} respectively.
Then, $\Delta$ is estimated from the CEF excitation energy in
PuO$_2$, experimentally found to be 123 meV.~\cite{Kern}
On the other hand, the Hund's rule coupling $J_{\rm H}$ between
$\Gamma_8$ and $\Gamma_7$ levels is 1/49 of the original Hund's
rule interaction among $f$ orbitals.~\cite{Hotta}
Namely, $J_{\rm H}$ is as large as a few hundred Kelvins.
Thus, we simply ignore the $\Gamma_7$ states in this paper.
As we will see later, this simplification is not
appropriate to reproduce experimental results quantitatively,
since the ground-state wave-function is not exactly reproduced.
However, we believe that our approximation provides a qualitatively
correct approach for the complex multipole state from the microscopic
viewpoint.

Since the $\Gamma_8$ quartet consists of two Kramers doublets,
we introduce `orbital' index $\tau$ ($=\alpha, \beta$)
to distinguish the two Kramers doublets,
while a `spin' index $\sigma$ ($=\uparrow, \downarrow$)
is defined to distinguish the two states in each Kramers doublet.
In the second-quantized form,
annihilation operators for $\Gamma_8$ electrons are given by
$c_{\bm{r} \alpha \uparrow}
=\sqrt{5/6} a_{\bm{r} 5/2}+\sqrt{1/6} a_{\bm{r} -3/2}$ and
$c_{\bm{r} \alpha \downarrow}
=\sqrt{5/6} a_{\bm{r} -5/2}+\sqrt{1/6} a_{\bm{r} 3/2}$
for $\alpha$-orbital electrons, and
$c_{\bm{r} \beta \uparrow}  =a_{\bm{r}  1/2}$ and
$c_{\bm{r} \beta \downarrow}=a_{\bm{r} -1/2}$
for $\beta$-orbital electrons,
where $a_{\bm{r} j_z}$ is
the annihilation operator for an electron
with the $z$-component $j_z$ of the total angular momentum
at site $\bm{r}$.

In the tight-binding approximation,
we obtain the following Hamiltonian for $\Gamma_8$ electrons:~\cite{Hotta}
\begin{eqnarray}
    \mathcal{H}
    &=&\sum_{\bm{r},\bm{\mu},\tau,\sigma,\tau^{\prime},\sigma^{\prime}}
    t^{\bm{\mu}}_{\tau \sigma; \tau^{\prime} \sigma^{\prime}}
    c^{\dagger}_{\bm{r} \tau \sigma}
    c_{\bm{r}+\bm{\mu} \tau^{\prime} \sigma^{\prime}}
    +U\sum_{\bm{r} \tau} n_{\bm{r} \tau \uparrow} n_{\bm{r} \tau \downarrow}
    \nonumber
    \\
    &+&U^{\prime}\sum_{\bm{r}} n_{\bm{r} \alpha} n_{\bm{r} \beta}
    +J\sum_{\bm{r},\sigma,\sigma^{\prime}}
    c^{\dagger}_{\bm{r} \alpha \sigma}
    c^{\dagger}_{\bm{r} \beta \sigma^{\prime}}
    c_{\bm{r} \alpha \sigma^{\prime}}
    c_{\bm{r} \beta \sigma}
    \nonumber
    \\
    &+&J^{\prime}\sum_{\bm{r},\tau \ne \tau^{\prime}}
    c^{\dagger}_{\bm{r} \tau \uparrow}
    c^{\dagger}_{\bm{r} \tau \downarrow}
    c_{\bm{r} \tau^{\prime} \downarrow}
    c_{\bm{r} \tau^{\prime} \uparrow},
\end{eqnarray}
where
$n_{\bm{r} \tau \sigma}$=$c^{\dagger}_{\bm{r} \tau \sigma} c_{\bm{r} \tau \sigma}$,
$n_{\bm{r} \tau}$=$\sum_{\sigma} n_{\bm{r} \tau \sigma}$,
$\bm{\mu}$ is a vector connecting nearest-neighbor sites,
and $t^{\bm{\mu}}_{\tau \sigma; \tau^{\prime} \sigma^{\prime}}$
is the hopping integral
of an electron with $(\tau^{\prime}, \sigma^{\prime})$
at site $\bm{r}$+$\bm{\mu}$
to the $(\tau, \sigma)$ state
at $\bm{r}$.
The coupling constants
$U$, $U^{\prime}$, $J$, and $J^{\prime}$
denote the intra-orbital Coulomb, inter-orbital Coulomb, exchange,
and pair-hopping interactions, respectively.

Concerning the hopping amplitudes,
we consider only
the overlap integrals of $f$-electron wave-functions
through the $\sigma$-bond $(ff\sigma)$.~\cite{Hotta,p-f}
For example, the hopping integrals between $f$-orbitals
at $(0,0,0)$ and $(a/2,a/2,0)$ ($a$ is the lattice constant) are given by
\begin{equation}
  t^{(a/2, a/2, 0)}_{\tau \uparrow; \tau^{\prime} \uparrow}=
  t^{(a/2, a/2, 0) *}_{\tau \downarrow; \tau^{\prime} \downarrow}=
  \begin{pmatrix}
    4 &  2\sqrt{3}i \\
    -2\sqrt{3}i & 3
  \end{pmatrix} t,
\end{equation}
and
\begin{equation}
  t^{(a/2, a/2, 0)}_{\tau \uparrow; \tau^{\prime} \downarrow}=
  t^{(a/2, a/2, 0)}_{\tau \downarrow; \tau^{\prime} \uparrow}= 0,
\end{equation}
where $t=(ff\sigma)/28$
and $t^{-\bm{\mu}}_{\tau \sigma; \tau^{\prime} \sigma^{\prime}}
=t^{\bm{\mu}}_{\tau \sigma; \tau^{\prime} \sigma^{\prime}}$.
Note that the hopping integrals depend on $\bm{\mu}$ and
they are intrinsically complex numbers in the fcc lattice.~\cite{comment}

In order to discuss multipole ordering,
we derive an effective model in the strong-coupling limit
using second-order perturbation theory with respect to $t$.
We consider the case of one electron per $f$ ion in the $\Gamma_8$
orbitals, but the effective model is the same for the one-hole case,
which corresponds to NpO$_2$, due to an electron-hole transformation.
Among the intermediate $f^2$-states in the perturbation theory,
we consider only the lowest-energy $\Gamma_5$ triplet states,
in which the two electrons occupy different orbitals,
assuming that other excited states are
well separated from the $f^2$ ground states.
In fact, the excitation energy from
the $\Gamma_5$ ground state of $f^2$ in UO$_2$
is 150~meV.~\cite{Amoretti}
Note that the CEF excitation energy is considered to be larger
than the triplet excitation one, since the Hund's rule interaction
is effectively reduced in the $j$-$j$ coupling scheme.~\cite{Hotta}
Thus, it is reasonable to take only the $\Gamma_5$ states
as the intermediate states.

\begingroup
\squeezetable
\begin{table*}
  \caption{
    \label{table:multipole_operators}
    Multipole operators in the $\Gamma_8$ subspace.~\cite{Shiina}
    The multipole operators are represented
    by pseudospin operators:
    $\hat{\bm{\tau}}=\sum_{\tau, \tau^{\prime}, \sigma}
    c^{\dagger}_{\tau \sigma}
    \bm{\sigma}_{\tau \tau^{\prime}}
    c_{\tau^\prime \sigma}$
    and
    $\hat{\bm{\sigma}}=\sum_{\tau, \sigma, \sigma^{\prime}}
    c^{\dagger}_{\tau \sigma}
    \bm{\sigma}_{\sigma \sigma^{\prime}}
    c_{\tau \sigma^\prime}$,
    where $\bm{\sigma}$ are the Pauli matrices.
    We use notations
    $\hat{\eta}^{\pm}=(\pm\sqrt{3} \hat{\tau}^x-\hat{\tau}^z)/2$
    and
    $\hat{\zeta}^{\pm}=-(\hat{\tau}^x \pm \sqrt{3}\hat{\tau}^z)/2$.
    The site label $\bm{r}$ is suppressed in this Table for simplicity.
  }
  \begin{ruledtabular}
    \begin{tabular}{c||c|cc|ccc|ccc|ccc|ccc}
      $\Gamma_{\gamma}$
      & $2u$   & $3gu$  & $3gv$
      & $4u1x$ & $4u1y$ & $4u1z$
      & $4u2x$ & $4u2y$ & $4u2z$
      & $5ux$  & $5uy$  & $5uz$
      & $5gx$  & $5gy$  & $5gz$ \\

      multipole operator $X^{\Gamma_{\gamma}}$
      & $T_{xyz}$   & $O^0_2$     & $O^2_2$
      & $J^{4u1}_x$ & $J^{4u1}_y$ & $J^{4u1}_z$
      & $J^{4u2}_x$ & $J^{4u2}_y$ & $J^{4u2}_z$
      & $T^{5u}_x$  & $T^{5u}_y$  & $T^{5u}_z$
      & $O_{yz}$    & $O_{zx}$    & $O_{xy}$ \\

      pseudospin representation
      & $\hat{\tau}^y$
      & $\hat{\tau}^z$
      & $\hat{\tau}^x$

      & $\hat{\sigma}^x$
      & $\hat{\sigma}^y$
      & $\hat{\sigma}^z$

      & $\hat{\eta}^+ \hat{\sigma}^x$
      & $\hat{\eta}^- \hat{\sigma}^y$
      & $\hat{\tau}^z \hat{\sigma}^z$

      & $\hat{\zeta}^+ \hat{\sigma}^x$
      & $\hat{\zeta}^- \hat{\sigma}^y$
      & $\hat{\tau}^x  \hat{\sigma}^z$

      & $\hat{\tau}^y \hat{\sigma}^x$
      & $\hat{\tau}^y \hat{\sigma}^y$
      & $\hat{\tau}^y \hat{\sigma}^z$ \\
    \end{tabular}
  \end{ruledtabular}
\end{table*}
\endgroup

After straightforward, but tedious calculations,
we arrive at an effective model in the form of
\begin{equation}
  \mathcal{H}_{\text{eff}}=
  \sum_{\bm{q}} (\mathcal{H}_{1 \bm{q}}
  +\mathcal{H}_{2 \bm{q}}+\mathcal{H}_{4u1 \bm{q}}+\mathcal{H}_{4u2 \bm{q}}),
  \label{eq:effectiveH}
\end{equation}
where $\bm{q}$ is the wave vector.
$\mathcal{H}_{1 \bm{q}}$ denotes
the interactions between quadrupole moments, given by
\begin{equation}
  \begin{split}
    \mathcal{H}_{1 \bm{q}}
    =a_1&(O^0_{2, -\bm{q}} O^0_{2, \bm{q}} c_x c_y + \text{c.p.})\\
    +a_3&(O^0_{2, -\bm{q}} O_{xy, \bm{q}} s_x s_y + \text{c.p.})\\
    +a_4&(O_{xy, -\bm{q}} O_{xy, \bm{q}} c_x c_y + \text{c.p.}),
  \end{split}
\end{equation}
where c.p. denotes cyclic permutations,
$c_{\nu}=\cos(q_{\nu} a/2)$,
and $s_{\nu}=\sin(q_{\nu} a/2)$ ($\nu=x$, $y$, or $z$).
The definitions of the multipole operators
and values of the coupling constants $a_i$ are
given in Tables~\ref{table:multipole_operators}
and \ref{table:coupling_constants}, respectively.
Note that
$O^0_{2 \bm{q}}$ transforms to $(\sqrt{3}O^2_{2 \bm{q}}-O^0_{2 \bm{q}})/2$
and $(-\sqrt{3}O^2_{2 \bm{q}}-O^0_{2 \bm{q}})/2$
under c.p. $(x,y,z)\rightarrow(y,z,x)$ and $(x,y,z)\rightarrow(z,x,y)$,
respectively.
$\mathcal{H}_{2 \bm{q}}$ and $\mathcal{H}_{4un \bm{q}}$ ($n=1$ or 2)
are the interactions between dipole and octupole moments, given by
\begin{equation}
  \begin{split}
    \mathcal{H}_{2 \bm{q}}
    =&b_8[T^{5u}_{z, -\bm{q}} T^{5u}_{z, \bm{q}} (c_y c_z + c_z c_x)
      + \text{c.p.}] \\
    +&b_9[T^{5u}_{x, -\bm{q}} T^{5u}_{y, \bm{q}} s_x s_y + \text{c.p.}] \\
    +&b_{10} T_{xyz, -\bm{q}} T_{xyz, \bm{q}} (c_x c_y + \text{c.p.}),
  \end{split}
\end{equation}
and
\begin{equation}
  \begin{split}
    \mathcal{H}_{4u n \bm{q}}
    =b^{(n)}_1&[J^{4u n}_{z -\bm{q}} J^{4u n}_{z \bm{q}} c_x c_y
      +\text{c.p.}] \\
    +b^{(n)}_2&[J^{4u n}_{z -\bm{q}} J^{4u n}_{z \bm{q}} (c_y c_z +c_z c_x)
      +\text{c.p.}] \\
    +b^{(n)}_3&[J^{4u n}_{x -\bm{q}} J^{4u n}_{y \bm{q}} s_x s_y+\text{c.p.}] \\
    +b^{(n)}_4&[T_{xyz -\bm{q}} (J^{4u n}_{z \bm{q}} s_x s_y+\text{c.p.})] \\
    +b^{(n)}_5&[T^{5u}_{z -\bm{q}} J^{4u n}_{z \bm{q}} c_z(c_x-c_y)
      +\text{c.p.})] \\
    +b^{(n)}_6&[T^{5u}_{z -\bm{q}}
      (-J^{4u n}_{x \bm{q}}s_z s_x+J^{4u n}_{y \bm{q}}s_y s_z) +\text{c.p.}],
    \label{eq:effectiveH_4u}
  \end{split}
\end{equation}
where values of the coupling constants $b_i$ and $b^{(n)}_i$
are shown in Table~\ref{table:coupling_constants}.
The above Eqs.~(\ref{eq:effectiveH})--(\ref{eq:effectiveH_4u})
are consistent with
the general form of multipole interactions on the fcc lattice
derived by Sakai~\textit{et al.}~\cite{Sakai}
We follow the notation in Ref.~\onlinecite{Sakai} for convenience.
\begin{table}
  \caption{\label{table:coupling_constants}
    Coupling constants in the effective model.
    The energy unit is $(1/16)t^2/(U^{\prime}-J)$.
  }
  \begin{ruledtabular}
    \begin{tabular}{ccccccccc}
      $a_1$       & $a_3$        & $a_4$ &
      $b_8$       & $b_9$        & $b_{10}$ &
      $b^{(1)}_1$ & $b^{(1)}_2$  & $b^{(1)}_3$ \\
      12          & $64\sqrt{3}$ & 192 &
      195         & $-336$       & 576 &
      $-196$      & $-4$         & 0           \\
      \hline
      $b^{(1)}_4$   & $b^{(1)}_5$ & $b^{(1)}_6$ &
      $b^{(2)}_1$   & $b^{(2)}_2$ & $b^{(2)}_3$ &
      $b^{(2)}_4$   & $b^{(2)}_5$ & $b^{(2)}_6$ \\
      $224\sqrt{3}$ & 0           & 0           &
      4             & 193         & $-336$      &
      $64\sqrt{3}$  & $2\sqrt{3}$ & $112\sqrt{3}$
    \end{tabular}
  \end{ruledtabular}
\end{table}

\begin{figure}
  \includegraphics[width=0.9\linewidth]{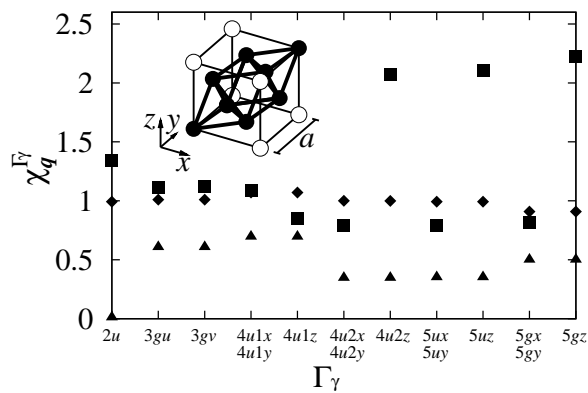}
  \caption{\label{figure:correlation_function}
    Correlation functions for the 8-site cluster for
    $\bm{q}=(0,0,0)$ (triangles),
    $\bm{q}=(0,0,1)$ (squares), and
    $\bm{q}=(1/2,1/2,1/2)$ (diamonds) in units of $2\pi/a$.
    The inset shows
    the fcc cluster (solid spheres) taken in the calculation.    
  }
\end{figure}

When we apply a mean-field theory to the effective model,
due care should be taken,
since in an fcc lattice with geometrical frustration,
the effect of fluctuations may be strong enough
to destroy the state obtained with mean-field theory.
Thus, we first evaluate
the correlation function in the ground state
using an unbiased method such as
exact diagonalization on the $N$-site lattice.
We set $N=8$, as shown in the inset of
Fig.~\ref{figure:correlation_function}.
The correlation function of the multipole operators is given by
$  \chi^{\Gamma_{\gamma}}_{\bm{q}}=(1/N)
  \sum_{\bm{r},\bm{r}^{\prime}} e^{i \bm{q} \cdot (\bm{r}-\bm{r}^{\prime})}
  \langle X^{\Gamma_{\gamma}}_{\bm{r}}
  X^{\Gamma_{\gamma}}_{\bm{r}^{\prime}} \rangle,$
where $\langle \cdots \rangle$ denotes
the expectation value
using the ground-state wave-function.
Figure~\ref{figure:correlation_function} shows
results for the correlation functions.
Although the interaction between $\Gamma_{2u}$ moments ($b_{10}$)
is large, the correlation function of the $\Gamma_{2u}$ moment
is not enhanced, indicating that
the frustration effect is significant for an Ising-like moment
such as $\Gamma_{2u}$.
Large values of correlation functions
are obtained for $J^{4u2}_{z}$, $T^{5u}_z$, and $O_{xy}$ moments
at $\bm{q}=(0,0,1)$ in units of $2\pi/a$.
We note that there is no term in the effective model
which stabilizes $O_{xy}$ quadrupole order at $\bm{q}=(0,0,1)$.
The enhancement of this correlation function indicates
an induced quadrupole moment
in $\Gamma_{4u2}$ or $\Gamma_{5u}$ moment ordered states.
Therefore, the relevant interactions are $b^{(2)}_2$ and $b_8$,
which stabilize the $J^{4u2}_z$ and $T^{5u}_z$ order, respectively,
at $\bm{q}=(0,0,1)$.
In the following, we consider a simplified model
including only $b^{(2)}_2$ and $b_8$.

Next we study the ordered state by applying mean-field theory
to our simplified model.
The coupling constant $b_8$ is slightly larger than $b^{(2)}_2$,
and $\Gamma_{5u}$ order has lower energy than $\Gamma_{4u2}$ order.
The interaction $b_8$ stabilizes longitudinal ordering of
the $\Gamma_{5u}$ moments,
but their directions are not entirely determined
by the form of the interaction.
In the $\Gamma_8$ subspace,
the $\Gamma_{5u}$ moment has an easy axis along [111].~\cite{Kubo,Kubo2}
Thus, taking the moment at each site
along [111] or other equivalent directions,
we find that a triple-$\bm{q}$ state is favored,
since it gains interaction energy in all the directions.
In fact, the ground state has longitudinal
triple-$\bm{q}$ $\Gamma_{5u}$ octupole order
with four sublattices, i.e.,
$(\langle T^{5u}_{x \bm{r}} \rangle,
\langle T^{5u}_{y \bm{r}} \rangle,
\langle T^{5u}_{z \bm{r}} \rangle) \propto
(\exp[i 2 \pi x/a],
\exp[i 2 \pi y/a],
\exp[i 2 \pi z/a])$.
Note that this triple-$\bm{q}$ structure
does not have frustration even in the fcc lattice.
The ground state energy is $-4b_8$ per site,
and the transition temperature is given by $k_{\text{B}} T_0=4b_8$.

\begin{figure}
  \includegraphics[width=0.95\linewidth]{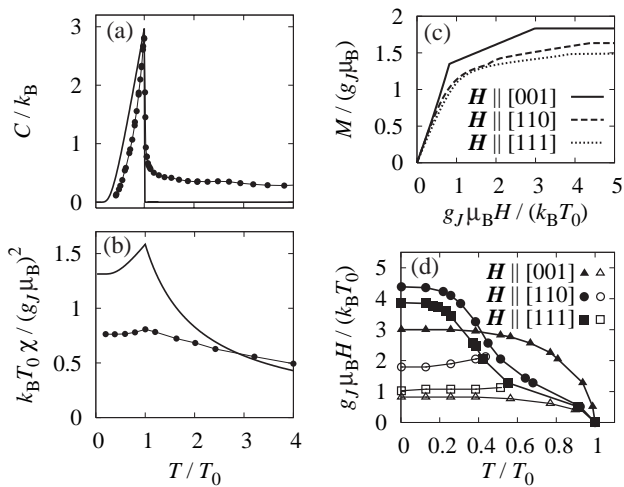}
  \caption{\label{figure:C_chi_M_and_PD}
    Results of the mean-field calculation.
    (a) Temperature dependence of the specific heat.
    Solid circles represent the experimental values~\cite{note:C}
    obtained by subtracting those for ThO$_2$~\cite{Osborne}
    from those for NpO$_2$.~\cite{Westrum}
    (b) Temperature dependence of the magnetic susceptibility.
    The Land\'{e} $g$-factor is $g_J=6/7$.
    Solid circles represent the experimental values
    for NpO$_2$.~\cite{Erdos}
    (c) Magnetic field dependence of the magnetization.
    (d) $H$-$T$ phase diagram.
    Solid symbols denote the $\Gamma_{5u}$ octupole transition,
    while open symbols denote
    transitions between
    $\Gamma_{5u}$ octupole ordered states
    with different sublattice structures.
    Note that $k_{\text{B}} T_0/(g_J \mu_{\text{B}})=43$~T for NpO$_2$.
  }
\end{figure}

Let us evaluate physical quantities in the mean-field theory.
Figures~\ref{figure:C_chi_M_and_PD}(a) and (b) show the temperature
dependence of the specific heat and magnetic susceptibility,
respectively.
The calculated results are compatible with
experimental ones for NpO$_2$, but
we should include higher energy states such as $\Gamma_7$
for quantitative agreement, as already mentioned.
This is one of future problems.
Figures~\ref{figure:C_chi_M_and_PD}(c) and (d) show
the magnetic field dependence of the magnetization at $T=0$
and an $H$-$T$ phase diagram, respectively.
The magnetization is isotropic as $H \rightarrow 0$
due to the cubic symmetry,
while anisotropy develops under a high magnetic field.
Note that under a high magnetic field,
sublattice structures change:
$(\langle T^{5u}_{x \bm{r}} \rangle,
\langle T^{5u}_{y \bm{r}} \rangle,
\langle T^{5u}_{z \bm{r}} \rangle) \propto
(0,0,\exp[i2\pi z/a])$   
for $\bm{H} \parallel [001]$,
a two-sublattice structure with
$\langle T^{5u}_{x \bm{r}} \rangle,
\langle T^{5u}_{y \bm{r}} \rangle \ne 0$,
and $\langle T^{5u}_{z \bm{r}} \rangle = 0$
for $\bm{H} \parallel [110]$,
and
$(\langle T^{5u}_{x \bm{r}} \rangle,
\langle T^{5u}_{y \bm{r}} \rangle,
\langle T^{5u}_{z \bm{r}} \rangle) \propto
(\sin[2\pi (y-z)/a],\sin[2\pi (z-x)/a],\sin[2\pi (x-y)/a])$
for $\bm{H} \parallel [111]$.
Note also that the triple-$\bm{q}$ state
is fragile under $\bm{H} \parallel [110]$:
$\langle T^{5u}_{z \bm{r}} \rangle=0$ with
a four-sublattice structure for
$g_J \mu_{\text{B}} H/(k_{\text{B}} T_0) \gtrsim 0.11$,
i.e., $H \gtrsim 5$~T at $T=0$
(this phase boundary is not shown
in Fig.~\ref{figure:C_chi_M_and_PD}(d)).

In order to confirm the octupole ordered state in NpO$_2$,
further experimental tests are required.
One possibility is a neutron diffraction measurement
under uniaxial pressure.
In an octupole ordered state,
a magnetic moment will be induced
by application of uniaxial pressure.~\cite{Kubo2,Kuramoto}
For example, a double-$\bm{q}$ magnetic moment
$(M_{x \bm{r}},M_{y \bm{r}},M_{z \bm{r}}) \propto
(\exp[i 2 \pi x/a],\exp[i 2 \pi y/a],0)$
is induced by pressure along the $z$-direction
in the longitudinal triple-$\bm{q}$ $\Gamma_{5u}$ octupole ordered state.

In summary, multipole order in $f$-electron systems on the fcc lattice
has been studied from a microscopic viewpoint.
An effective model has been derived
from a microscopic model based on the $j$-$j$ coupling scheme.
Multipole correlation functions in the ground state
of the effective model have been calculated
by exact diagonalization.
It has been revealed that the interactions for $\Gamma_{4u2}$
and $\Gamma_{5u}$ moments are dominant.
Mean-field theory has been applied to the simplified model
including only such interactions,
and longitudinal triple-$\bm{q}$ $\Gamma_{5u}$ octupole order
has been found to be realized.

We thank S. Kambe, N. Metoki, H. Onishi, Y. Tokunaga, K. Ueda,
R. E. Walstedt, and H. Yasuoka for discussions.
One of the authors (T. H.) is supported by Grant-in-Aids for Scientific
Research of Japan Society for the Promotion of Science
and of the Ministry of Education, Culture, Sports,
Science, and Technology of Japan.


\begin{thebibliography}{99}
\bibitem{p-f}
Even if we include the hopping via oxygen $p$ orbitals,
the hopping matrix is written in the form of Eq.~(2).
\bibitem{comment}
It is in sharp contrast to a model on a simple cubic lattice,~\cite{Hotta}
in which the hopping integrals are real and
the same form as for $e_g$ orbitals of $d$ electrons.
In other words, there is no difference
between the $\Gamma_8$ model for $f$ electrons
and the $e_g$ orbital model for $d$ electrons
on the simple cubic lattice
in the absence of a magnetic field.
For $e_g$ orbitals, we can always set the hopping integrals to be real,
irrespective of the lattice type,
by selecting appropriate basis wave-functions,
while $\Gamma_8$ orbitals on the fcc lattice
appear to be complex in nature,
specific to $f$-electron systems
with strong spin-orbit coupling.
\bibitem{note:C}
To subtract specific heat values of ThO$_2$~\cite{Osborne}
from those of NpO$_2$,~\cite{Westrum}
we interpolate the experimental data of ThO$_2$
and evaluate the values at temperatures in which specific heat
of NpO$_2$ have been measured.
\end{thebibliography}
\end{document}